\documentclass[prl,twocolumn]{revtex4}
\usepackage{graphicx}
\usepackage{amsmath,amsbsy,amssymb}
\usepackage{bm}
\usepackage{mathrsfs,color}

\newcommand{\diff}{\mathrm{d}}

\newcommand{\imu}{\mathrm{i}}
\newcommand{\epn}{\mathrm{e}}

\newcommand{\ua}{\uparrow}
\newcommand{\da}{\downarrow}
\newcommand{\dg}{\dagger}
\newcommand{\la}{\langle}
\newcommand{\ra}{\rangle}
\newcommand{\al}{\alpha}
\newcommand{\sg}{\sigma}
\newcommand{\gm}{\gamma}
\newcommand{\ep}{\varepsilon}

\begin{document}

\title{
Superconductivity from emerging magnetic moments
}

\author{Shintaro Hoshino$^{1,2}$ and Philipp Werner$^2$}

\affiliation{
$^1$Department of Basic Science, The University of Tokyo, Meguro, Tokyo 153-8902, Japan
\\
$^2$Department of Physics, University of Fribourg, 1700 Fribourg, Switzerland
}

\date{\today}

\begin{abstract}
Multi-orbital Hubbard models are shown to exhibit a spatially isotropic spin-triplet superconducting phase, where equal-spin electrons in different local orbitals are paired. This superconducting state is stabilized in the spin-freezing crossover regime, where local moments emerge in the metal phase, and the pairing is substantially assisted by spin anisotropy. The 
phase diagram features a superconducting dome below a non-Fermi liquid metallic region and next to a magnetically ordered phase.  We suggest that this type of fluctuating-moment induced superconductivity,
which is not originating from fluctuations near a quantum critical point, 
may be realized in spin-triplet superconductors such as strontium ruthenates 
and uranium compounds.
\end{abstract}

\maketitle

Spin-triplet superconductivity, 
in the sense of equal-spin pairing, 
is believed to occur in a number of correlated materials. 
The best candidate is the layered compound Sr$_2$RuO$_4$, where the Knight shift 
remains unchanged 
across the superconducting phase boundary, in stark contrast with the behavior expected for spin-singlet pairing \cite{maeno2012}. 
In the iron pnictides, where a spin-triplet superconducting phase has been proposed in early theoretical works \cite{dai2008}, the experimental evidence points toward spin-singlet pairing, although in LiFeAs a spin-triplet scenario is still being debated \cite{baek2013,brand2014}.
The uranium based superconductors are also possible candidates for spin-triplet pairing. In compounds such as UGe$_2$, UCoGe and URhGe, the superconducting state is found near a ferromagnetic phase and the two orders may even coexist \cite{saxena2000, aoki2001, huy2007, aoki2012}.
For a deeper understanding of unconventional superconductivity in strongly correlated electron systems with multiple active orbitals, it is thus important to clarify the mechanisms which can lead to spin-triplet pairing.

\begin{figure*}[ht]
\begin{center}
\includegraphics[width=179mm]{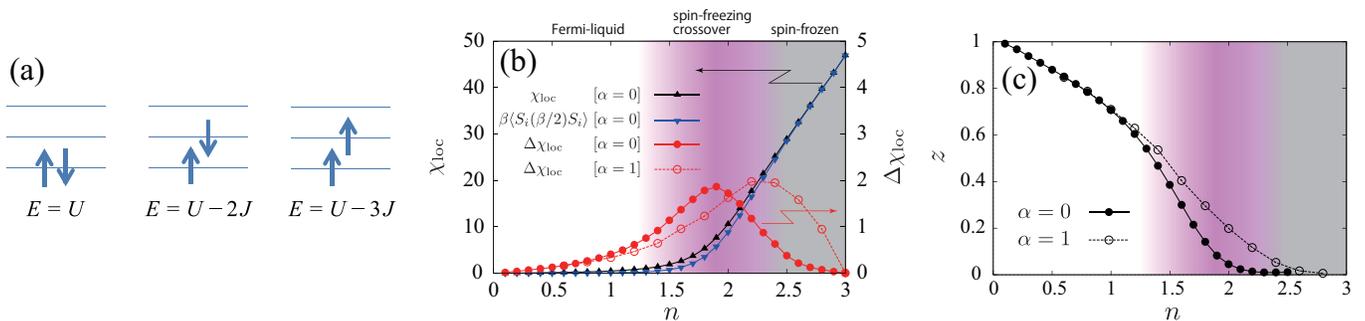}
\caption{
(a) Illustration of possible local configurations with two electrons in three orbitals, and the corresponding energies $E$.  
(b) Filling dependence of the local magnetic susceptibilities for $U=1$ and $T=0.005$.
The black curve shows the local magnetic susceptibility $\chi_\text{loc}$, and the blue curve the contribution 
from 
long-lived (frozen) moments. 
(c) Filling dependence of the renormalization factor $z$ for the same parameters. 
}
\label{fig:concept}
\end{center}
\end{figure*}

While a $p$-wave symmetry is usually assumed for the pairing state in spin-triplet superconductors, 
an
$s$-wave spin-triplet pairing 
is 
also possible by taking into account the orbital degrees of freedom. 
The mechanism of this unconventional superconductivity can be easily understood 
\cite{klejnberg1999, spalek2001, han2004,sakai2004,kubo2007,dai2008,zegrodnik2013,zegrodnik2014}: 
same-spin electrons tend to occupy the same site due to the Hund coupling 
which favors high-spin states. 
A new insight in this paper is that the $s$-wave spin-triplet pairing is closely connected to the emergence of local magnetic moments 
in 
so-called Hund metals \cite{georges2013}. 
This class of materials, which includes ruthenates \cite{werner2008, medici2011} and iron pnictides \cite{haule2009, liebsch2010,werner2012,yin2012,toschi2012,liu2012}, exhibits Hund-coupling induced correlation effects and characteristic non-Fermi liquid properties. 
The underlying phenomenon is {\it spin-freezing} \cite{werner2008}: In a 
narrowly defined range of fillings and interaction strengths, long-lived magnetic moments appear in the metal phase
of multi-band systems 
with Hund coupling (formation of a large composite spin, see right panel of Fig.~\ref{fig:concept}(a)).
In the absence of long-range order, the emerging local moments will be screened at sufficiently low temperature, so that there is no quantum phase transition associated with 
spin freezing. 
However, screening large local moments is difficult, and the
Fermi liquid coherence temperature becomes very low \cite{georges2013,nevidomskyy2009}.
Hence, as demonstrated here,  
a spontaneous symmetry breaking pre-empts the 
screening of the moments.
While deep in the spin-frozen regime
the long-lived local moments order magnetically at low temperatures, 
the emerging and 
fluctuating 
local moments in the spin-freezing crossover regime generate spin-triplet pairing.
This leads to the formation of a superconducting dome 
separating the Fermi liquid metal from 
the magnetically ordered region
and results in phase diagrams which
closely resemble 
those of 
unconventional superconductors.

We consider a 
three-orbital
 Hubbard model whose Hamiltonian is given by
\begin{align}
&\mathscr{H} = \sum_{\bm k\gm\sg} (\ep_{\bm k}-\mu) c^\dg_{\bm k\gm\sg} c_{\bm k\gm \sg}
+ U \sum_{i\gm} n_{i\gm\ua} n_{i\gm\da} \label{hamilt} \\
&
+ U' \sum_{i\sg,\gm<\gm'} n_{i\gm\sg} n_{i\gm'\bar \sg}
+ (U'-J) \sum_{i\sg,\gm<\gm'} n_{i\gm\sg} n_{i\gm'\sg}
\nonumber \\
& 
- \alpha J \sum_{i,\gm<\gm'} (
 c^\dg_{i\gm\ua} c_{i\gm\da} c^\dg_{i\gm'\da} c_{i\gm'\ua}
+c^\dg_{i\gm\ua} c^\dg_{i\gm\da} c_{i\gm'\ua} c_{i\gm'\da}
+{\rm H.c.} ), \nonumber
\end{align}
where $i$ is the site index, 
$\gm=1,2,3$
the orbital index, $\sigma=\ua,\da$ the spin index, and  
$\bar \sg$ represents the complementary spin ($\bar \ua \!\!= \da$).
$\ep_{\bm k}$ is the dispersion of electrons on the lattice, and $\mu$ is the chemical potential.
The interaction terms contain the intra-orbital ($U$) and inter-orbital ($U'$) Coulomb repulsions, and the Hund coupling $J$. 
The parameter $\al$ controls the anisotropy in spin space, i.e. $\alpha=1$ corresponds to a spin-rotationally invariant system and $\alpha=0$ to the Ising anisotropic case where the interactions are only of density-density type.
A spin anisotropy may originate from 
spin-orbit coupling, and the parameter $\alpha$ allows us to incorporate this effect in a simple manner 
\cite{footnote_soc}.

An anisotropic coupling in spin space should in principle only change the prefactor of the spin-flip term in Eq.~\eqref{hamilt}. 
However, for $J>0$ the pair-hopping term, which transfers two electrons in the same orbital to another orbital, 
is not important.
This is because the pair-hopping favors the state shown in the 
left
panel of Fig.~\ref{fig:concept}(a),
which is hardly 
realized due to the presence of the intra-site Coulomb interaction $U$ which is larger than $J$.
We therefore consider it more convenient to put the anisotropy factor in front of both terms, so that $\alpha$ interpolates between the familiar Ising and rotationally invariant limits.

While the phenomena discussed in this paper are 
generic features of multi-orbital systems with non-zero Hund coupling parameter $J$, we will show results for the 3-orbital case 
with 
$U'=U-2J$, $J/U=1/4$  and consider a semi-circular density of states with bandwidth $W=1$. 
(We neglect specific
material effects related to the particular shape of the density of states and to the
hybridization between different orbitals, aiming at a general,
material unspecific, discussion of the physics.)
The model is solved using the dynamical mean field theory (DMFT) \cite{georges1996}, combined with a numerically exact continuous-time Monte Carlo method \cite{werner2006matrix, gull2011}. 
This formalism 
captures 
local correlation effects.

To illustrate the spin-freezing phenomenon, 
we compute the 
dynamic contribution to the 
local magnetic susceptibility 
\begin{align}
\Delta \chi_{\rm loc} = \int_0^\beta \diff \tau \left(
\la S_i(\tau) S_i \ra - \la S_i(\beta/2) S_i \ra
\right)
, \label{eq:loc_chi}
\end{align}
where $\beta=1/T$ is the inverse temperature and ${\mathscr O}(\tau) = \epn^{\tau {\mathscr H}} {\mathscr O} \epn^{-\tau {\mathscr{H}}} $.
The operator $S_i = \tfrac 1 {2M} \sum_{\gm=1}^M (c^\dg_{i\gm \ua}c_{i\gm \ua}-c^\dg_{i\gm \da}c_{i\gm \da})$, with $M=3$ the number of orbitals, measures the local spin.
The first term on the right-hand side yields the local magnetic susceptibility ($\chi_{\rm loc}$). 
In Eq.~\eqref{eq:loc_chi} we subtract the long-time correlator $\la S_i(\beta/2) S_i \ra$, which reflects the magnitude of long-lived frozen moments \cite{werner2008,hafermann2012}. 
Hence, the quantity $\Delta \chi_{\rm loc}$ measures the fluctuations of the moments.
Figure~\ref{fig:concept}(b) shows the filling dependence of these quantities for $\al=0$. While the local susceptibility $\chi_{\rm loc}$ monotonically increases with increasing $n$, the fluctuation $\Delta \chi_{\rm loc}$ reaches a maximum at $n\simeq 1.9$.
This peak 
indicates the crossover between the Fermi liquid and spin-frozen regimes, and we use the location of the maximum as our definition of the ``spin-freezing crossover."
The spin-freezing 
is also reflected in the renormalization factor $z$, or mass-enhancement factor $1/z$, of the quasi-particles \cite{medici2011}.
For the estimation,
we use the ansatz $\Sigma(\omega\rightarrow 0) = a+b\omega$ and determine the coefficients by fitting the numerical data. 
Specifically, we fit the self energy by the Pad\'{e} approximation using the lowest two Matsubara frequencies, and compute the renormalization factor by the relation $z = (1-b)^{-1}$.
Figure~\ref{fig:concept}(c) exhibits a drop of $z$ in the spin-freezing crossover region. 
Qualitatively similar results are obtained in the SU(2) symmetric case ($\alpha=1$, dashed lines).

\begin{figure*}[t]
\begin{center}
\includegraphics[width=179mm]{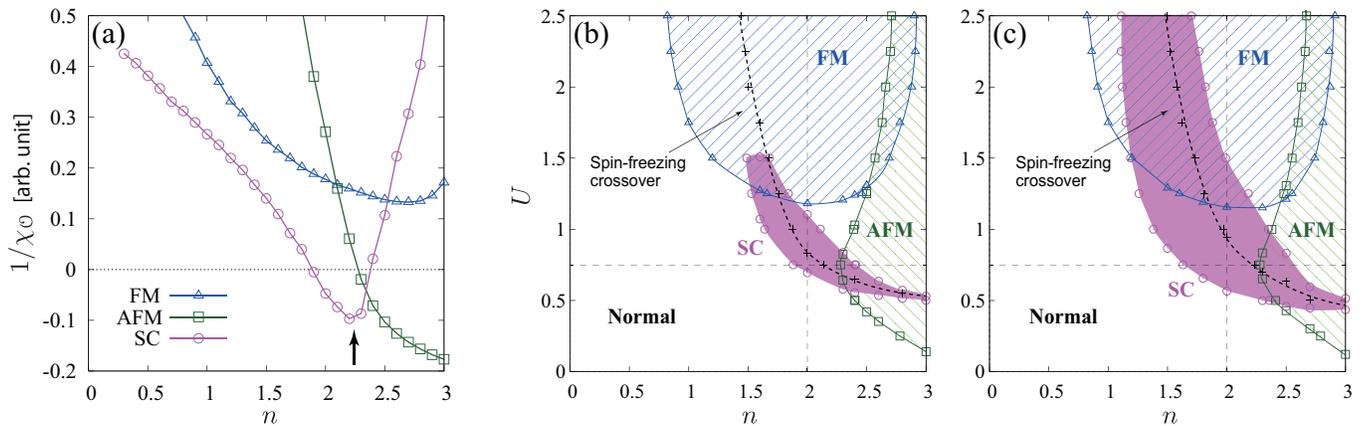}
\caption{
(a) Filling dependence of the inverse susceptibilities for FM, AFM and SC orders for $U=0.75$, $T=0.005$ in the system with Ising spin anisotropy ($\alpha=0$).
A negative $1/\chi_\mathcal{O}$ indicates a long-range ordered phase with order parameter $\mathcal{O}$. 
(b,c) Interaction-filling phase diagrams at $T=0.005$ and $T=0.0025$, respectively.
The black dashed line shows the location of the spin-freezing crossover in the system without long-range order. 
}
\label{fig:phase}
\end{center}
\end{figure*}

To study the stability regions of ordered phases, we calculated the susceptibilities 
\begin{align}
\chi_{\mathscr O} = \frac 1 N \int_0^{\beta} \la {\mathscr O}(\tau) {\mathscr O}^\dg \ra \, \diff \tau
,
\end{align}
where $N$ is the total number of sites.
The operator ${\mathscr O}$ is given by
\begin{align}
{\mathscr O} = \left\{
\begin{matrix}
\sum_{i} S_i & ({\rm FM}) \\
\sum_i \lambda_i S_i & ({\rm AFM}) \\
\sum_{i} c^\dg_{i\gm\ua} c^\dg_{i\gm'\ua} {\rm \ \ for\ } \gm\neq\gm'  & ({\rm SC}) \\
\end{matrix}
\right.
\label{order_param}
\end{align}
for ferromagnetic order (FM), anti-ferromagnetic order (AFM) and $s$-wave inter-orbital spin-triplet superconductivity (SC). 
$\lambda_i$ is a sign 
which depends on the sublattice.
A divergence in $\chi_{\mathscr O}$ (or equivalently a sign-change in $1/\chi_{\mathscr O}$) indicates a possible transition into  
a long-range ordered 
phase. 
The susceptibilities in Eq.~\eqref{order_param} can be derived from 
the two-particle Green's function 
\cite{supp}.  
We calculate the vertex part of this Green's function from the local impurity problem, and obtain the lattice two-particle Green's function by solving the Bethe-Salpeter equation.
While we have also calculated the susceptibilities for other types of orders, such as orbital ordering, only the quantities listed in Eq.~\eqref{order_param} diverge in the parameter regions considered in this paper.

Figure~\ref{fig:phase}(a) shows the inverse susceptibilities 
for $T=0.005$, $U=0.75$, $\alpha=0$ and different fillings. 
Symmetry broken phases exist in the regions where $1/\chi_{\mathscr O}<0$.
Repeating this analysis for different $U$,  
we obtain the $T=0.005$ phase diagram 
shown in Fig.~\ref{fig:phase}(b). 
At $U\gtrsim 1.25$,
a FM phase appears \cite{chan2009}, while near half-filling the AFM phase is stable. 
A new result is the existence of 
a 
SC region connecting the FM and AFM phases. 
This spin-triplet SC phase is clearly associated with the spin-freezing crossover, indicated by the black dashed line, 
which suggests 
that the fluctuating local moments at the border of the spin-frozen regime induce 
the 
pairing.
We also show the phase diagram at a lower $T=0.0025$ in Fig.~\ref{fig:phase}(c), where the SC region expands.

We note that there is no direct attraction among electrons, although the superconductivity is realized by forming electron pairs.
The effective attraction for Cooper pairs can be understood from the imbalance of the Coulomb interactions \cite{inaba2012, zegrodnik2013, 
okanami2014, 
koga2014}.
To see this, let us consider the situation with two electrons on one lattice site.
As shown in Fig.~\ref{fig:concept}(a), there are three kinds of configurations.
The energetically most favorable one is the electron pair with the same spins, because of the Hund coupling $J$.
Hence, two same spin electrons will tend to occupy the same site (but different orbitals), even though a repulsive Coulomb interaction acts between them.
Indeed, the local effective interaction $\tilde U$ among same-spin electrons can be derived within second-order perturbation theory \cite{inaba2012}, which results in
\begin{align}
\tilde U \simeq
(U'-J) - [2UU' + (U'-J)^2 + U'^2] \chi_{\rm loc},
\end{align}
with $\chi_{\rm loc}=\Delta \chi_{\rm loc}$ in the weak-coupling approximation \cite{supp}.
If the second term on the right-hand side dominates the first-order term due to a large $\chi_{\rm loc}$, the interaction becomes attractive. 

A reduction of the interaction energy below $U-3J$ 
is not possible if the number of electrons is constrained to two per site, on average.
On the other hand, the electrons do not have to occupy the same site when $n\leq 1$, and the superconducting state is never realized in this case.

\begin{figure}[t]
\begin{center}
\includegraphics[width=\columnwidth]{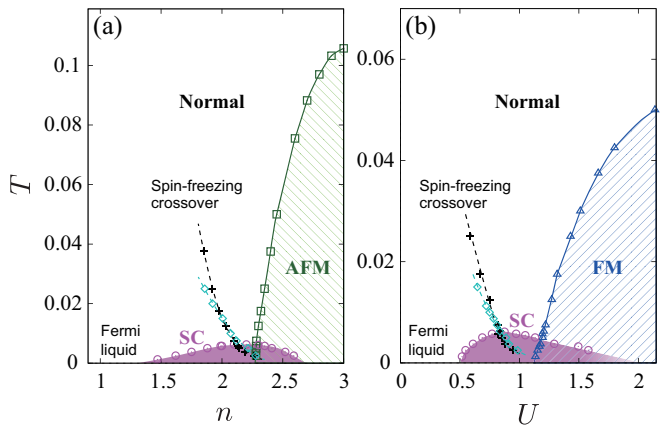}
\caption{
(a) Temperature-filling phase diagram at $U=0.75$.
(b) Temperature-interaction phase diagram at $n=2$.
The black dashed line indicates the spin-freezing crossover in the system with suppressed long-range order. 
The light blue diamonds show the minimum of the inverse pairing susceptibility. 
}
\label{fig:phase_temp}
\end{center}
\end{figure}

Next, we discuss the temperature-filling phase diagram shown for $U=0.75$ in Fig.~\ref{fig:phase_temp}(a). 
With hole doping from half-filling ($n=3$), the AFM transition temperature decreases and 
becomes zero 
at $n\simeq 2.3$, which is close to the spin-freezing line. 
By further doping with holes, we find the spin-triplet 
SC 
phase with a dome-shaped $T_c$. 
If we fix the filling to $n=2$ and change $U$, we obtain the phase diagram shown in Fig.~\ref{fig:phase_temp}(b).
With decreasing $U$ (increasing pressure, experimentally), the FM order is destroyed and again a 
SC 
dome appears 
next to the magnetic region. 
The light blue diamonds in Fig.~\ref{fig:phase_temp} 
show 
the points where the inverse pairing susceptibility reaches its minimum 
as indicated by the arrow in Fig.~\ref{fig:phase}(a). Above $T_c$, this corresponds to the maximal pairing instability, while 
below $T_c$ this minimum approximately locates the maximum order parameter \cite{supp}. 
The close resemblance 
 with the spin-freezing crossover line shows the relevance of spin freezing for the present superconductivity.

SC 
domes are usually understood as a manifestation of fluctuations associated with 
magnetic
quantum critical points. 
However, the superconductivity revealed in this paper is induced by {\it local} magnetic fluctuations in the spin-freezing crossover regime. 
Nevertheless, the 
SC
order naturally appears in the vicinity of a magnetic phase, since the strengthening of the magnetic moments 
deeper inside the spin-frozen regime causes magnetic ordering. 
Furthermore, the normal state above the 
SC 
dome is a non-Fermi liquid 
whose properties are influenced by 
the spin-freezing crossover \cite{werner2008}.

\begin{figure}[t]
\begin{center}
\includegraphics[width=60mm]{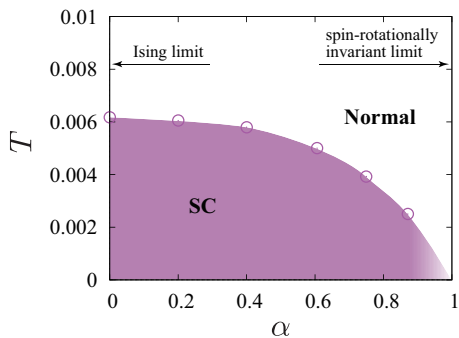}
\caption{
Dependence of the superconducting transition temperature on the spin anisotropy $\alpha$ for $U=0.875$ and $n=2$.
}
\label{fig:alpha_depend}
\end{center}
\end{figure}

So far we have  
shown results for 
the system with Ising anisotropy.
We now clarify how the superconductivity is affected by the spin-flip term 
in the model with $\al\neq 0$.
Figure \ref{fig:alpha_depend} shows 
$T_c$ 
for
$U=0.875$ and filling $n=2$.
As $\alpha$ is increased from $0$, the transition temperature decreases and 
drops below the lowest accessible temperature at $\al=1$.
The destabilization of the electron pairs by the spin-flip term can be intuitively understood by looking at Fig.~\ref{fig:concept}(a).
Since the spin-flip term exchanges $\ua$ and $\da$ spin electrons residing in different orbitals, 
the configuration shown in the middle panel is favored.
As a result, the 
probability for the equal-spin state (right panel) decreases 
compared to the spin-anisotropic case.
Near $\alpha=1$, the fluctuations among the 3 degenerate spin-triplet states further destabilize the equal-spin pairing.
In fact, a previous study of a single-orbital model based on the Eliashberg theory \cite{monthoux1999} demonstrated the importance of {\it longitudinal} fluctuations for pairing, which is consistent with our results.

Finally, let us comment on the potential
implications 
of these findings for unconventional multi-band superconductors. 
Because an Ising-type spin-freezing is underlying the fluctuating-moment induced spin-triplet superconductivity, it may be expected to occur in electron systems with strong 
spin-orbit coupling.
Promising candidates are the uranium-based superconductors UGe$_2$ \cite{saxena2000}, URhGe \cite{aoki2001} and UCoGe \cite{huy2007}, which exhibit a 
SC 
phase bordering a FM phase.
In these compounds a strong Ising spin-anisotropy is observed: the magnetization along the easy axis is several times larger than along the hard axis \cite{aoki2012}.
Futhermore, Ising-type spin-fluctuations are important for the superconductivity, since a magnetic field along the magnetic moment destabilizes the pairing, while it is much more robust against fields perpendicular to the moment \cite{aoki2012,hattori2012}.
The 5$f$ electrons in the 
U 
ions, which play a central role in the low-temperature behavior,  
have a relatively itinerant nature and are strongly correlated.
Although 
a realistic description should involve an 
Anderson lattice, the local interaction is of the Slater-Kanamori type, and hence we expect the same spin-triplet 
SC state. 
We thus believe that our mechanism 
could be realized in 
these uranium-based superconductors.

Sr$_2$RuO$_4$ is another candidate compound which 
might exhibit 
a fluctuating-moment induced 
superconductivity. 
Here, the spin-orbit coupling is nearly 100 meV \cite{maeno2012}, and as shown in Fig.~\ref{fig:alpha_depend} the spin anisotropy need not be very large to realize 
a SC state 
at low temperatures. 
Also, the estimated $U\simeq 0.8$ \cite{medici2011} and the filling $n=4$ (same as $n=2$ due to particle-hole symmetry) place this material in the parameter regime where 
the SC state 
is found near 
a FM 
phase (Figures~\ref{fig:phase}(b) and (c)). 
A related compound, SrRuO$_3$, with a larger $U$,
becomes a ferromagnet \cite{georges2013} and
exhibits the non-Fermi liquid behavior associated with spin freezing in the 
high-temperature 
phase \cite{kostic1998, dodge2000}.

In the iron pnictides, the Coulomb interactions and fillings on the Fe sites can also be 
close to the spin-freezing crossover values 
\cite{liebsch2010,werner2012}, and for LiFeAs, in particular, the experimental signatures fully support this interpretation \cite{wright2013}. 
On the other hand, the 3$d$ electrons have weak spin-orbit coupling, 
and hence a small spin anisotropy.
Thus, for iron pnictides (including LiFeAs),
one can expect  other types of pairing driven by e.g. Fermi-surface nesting mechanisms to dominate.

For an experimental detection of the present superconductivity, it is necessary to measure both the spin and spatial parts 
of the pairing state. 
The spin-part can be determined by measuring the magnetic susceptibility, e.g. with NMR. If it is identified as spin-triplet, one still has to distinguish between $p$-wave pairing and the proposed $s$-wave inter-orbital pairing. The difference lies in the presence/absence of a node in the gap function which will be reflected in a power-law/exponential temperature dependence of physical quantities.
Our work provides a general 
guiding principle in the search for new unconventional multi-band superconductors, namely the combination of emerging local moments in the spin freezing crossover regime  
and 
spin anisotropy
in heavy elements.

%
{\it Acknowledgements.} 
We thank Ch. Bernhard, A. Koga, A. Millis and M. Sigrist for helpful discussions. 
S.H. acknowledges financial support 
from JSPS KAKENHI Grant No. 13J07701 and P.W. support from FP7 ERC starting grant No. 278023.
The authors benefited from the Japan-Swiss Young Researcher Exchange Program 2014 coordinated by JSPS and SERI.
The numerical calculations have been performed on the BEO04 cluster at the University of Fribourg and the supercomputer at ISSP (University of Tokyo).

\clearpage
\onecolumngrid

\setcounter{section}{0}
\setcounter{equation}{0}
\setcounter{figure}{0}

\vspace{5mm}
\section*{\large
Supplementary material for\\
``Superconductivity from emerging magnetic moments''
}
\vspace{5mm}

\section*{Effective attraction from purely repulsive interactions}

Here, we explain how local fluctuations can induce a pairing among repulsively interacting electrons. 
We follow Ref.~1 that deals with a three-component fermion system.
In the weak-coupling regime, the effective interactions that incorporate bubble diagrams can be written in general form as
\begin{align}
\tilde U_{\al\beta} (q) = U_{\al\beta} - \sum_{\alpha_1} U_{\al\al_1} \chi_{\alpha_1}(q) \tilde U_{\alpha_1\beta} (q)
, \label{eq:eff_int}
\end{align}
where $q = (\bm q, \imu \nu_m)$ with $\bm q$ denoting the wave vector and $\nu_m=2\pi m T$ a bosonic Matsubara frequency.
For the present three orbital Hubbard model, the indices are given by $\alpha = (\gm,\sg)$ where $\gm = 1,2,3$ and $\sg=\ua,\da$.
The bare interactions are given by $U_{\gm\sg\mathchar`-\gm\sg}=0$, $U_{\gm\ua\mathchar`-\gm\da}=U$, $U_{\gm\ua\mathchar`-\gm'\da}=U'$, $U_{\gm\ua\mathchar`-\gm'\ua}=U'-J$ ($\gm \neq \gm'$).
The dynamical susceptibility is defined by $\chi_\alpha (q) = - \sum_k g_\al(k)g_\al(k+q)$ where $g_\alpha (k)$ is the single-particle Green function for electrons with flavor $\alpha$.
For the case of degenerate orbitals considered in our paper, we do not need the index $\alpha$ in the susceptibility.

\begin{figure*}[b]
\begin{center}
\includegraphics[width=150mm]{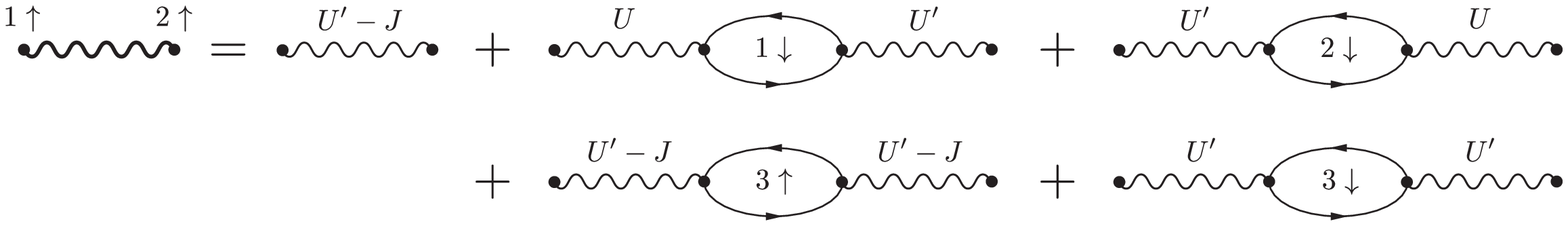}
\caption{
Effective attractive interactions from bubble diagrams up to second order.
}
\label{fig:diagram}
\end{center}
\end{figure*}

In the 
DMFT 
approximation, only the local part of the vertex corrections is taken into account [2].
Hence we replace the susceptibility by the local one, $\chi_{\rm loc} (\imu \nu_m)$.
(This replacement is not essential for the pairing: the effective attraction can be derived even when we consider the $\bm q$-dependent susceptibility, as discussed in Ref.~1.)
By solving Eq.~\eqref{eq:eff_int}, the static interaction among $1\ua$ and $2\ua$ electrons can be explicitly derived as
\begin{align}
\tilde U_{1\ua\mathchar`-2\ua} (0) = 
\frac{
U'-J + (J^2 - 2UU' - 2U'J) \chi_{\rm loc} + (U'-J)(U^2 - 2J^2 + 4U'J)\chi_{\rm loc}^2
}{
[1-(U-J)\chi_{\rm loc}]
[1-(U+2J)\chi_{\rm loc}]
[1+(U-2U'+J)\chi_{\rm loc}]
[1+(U+4U'-2J)\chi_{\rm loc}]
}
,
\end{align}
where we consider the static component: $\chi_{\rm loc} = \chi_{\rm loc}(0)$.
The diagrams up to second-order in the interactions are shown in Fig.~\ref{fig:diagram}. In this approximation the effective interaction is given by
\begin{align}
\tilde U_{1\ua\mathchar`-2\ua} (0) \simeq
U'-J - [2UU' + (U'-J)^2 + U'^2] \chi_{\rm loc} 
. \label{eq:second-order}
\end{align}
Thus if the second-order terms dominate the bare interaction $U'-J$, the effective interaction $\tilde U_{1\ua\mathchar`-2\ua}$ can become attractive even though the bare interaction is repulsive. 
Hence, Eq.~\eqref{eq:second-order} shows that strong 
local fluctuations induce a pairing among electrons.
This argument is valid in the case of  
weak interactions, where no local moments are formed.
In this regime, the relation $\Delta\chi_{\rm loc}=\chi_{\rm loc}$ holds  
(see Eq. (2) in the main text for the definition of $\Delta\chi_{\rm loc}$).

In the above argument, the
local susceptibility can be identified as the magnetic and charge susceptibilities, which have the same value in the weak-coupling limit. 
With increasing repulsive Coulomb interactions, the magnetic susceptibility is enhanced and the charge one is suppressed. 
Hence we expect that in the regime considered in the main text, the local magnetic fluctuations primarily contribute to the pairing among electrons in the multi-orbital Hubbard model.
Indeed our DMFT+CTQMC calculations demonstrate a clear connection between superconductivity and local spin susceptibility.
We note that the present discussion cannot be applied to the local-moment regime with $\Delta\chi_{\rm loc} \neq \chi_{\rm loc}$.
In this case the expansion from the strong-coupling limit should work as an effective theory.

\begin{figure*}[t]
\begin{center}
\includegraphics[width=130mm]{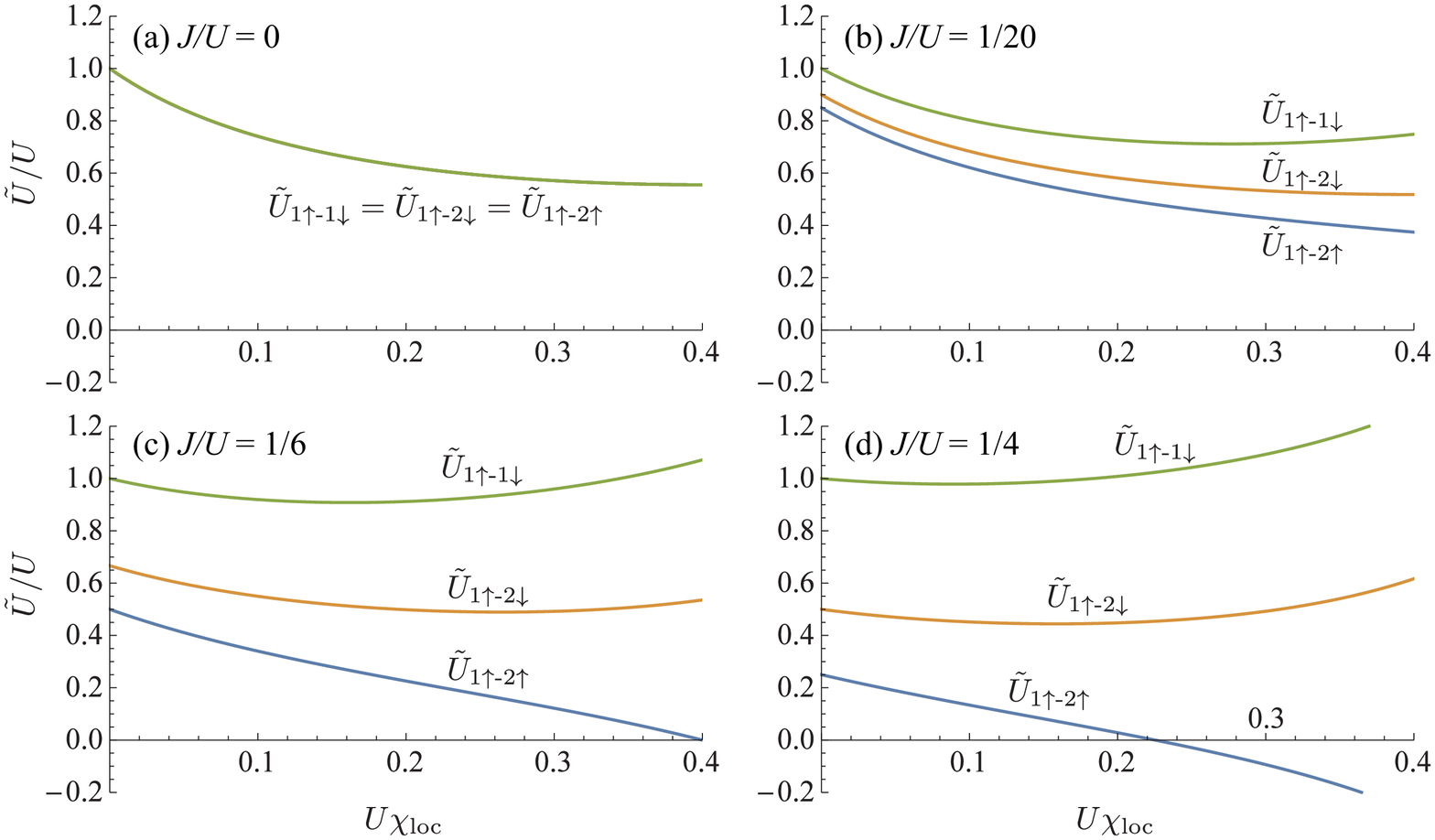}
\caption{
Effective local interactions as a function of the local susceptibility.
}
\label{fig:interaction}
\end{center}
\end{figure*}

The other effective interactions can be derived in a similar manner.
Figure~\ref{fig:interaction} shows the local-fluctuation (susceptibility) dependence of the effective interactions for several values of $J/U$.
Note that we have used the relation $U'=U-2J$.
Here we have only the three independent components $\tilde U_{1\ua\mathchar`-1\da}$, $\tilde U_{1\ua\mathchar`-2\da}$ and $\tilde U_{1\ua\mathchar`-2\ua}$.
As shown in Fig.~\ref{fig:interaction}(a,b), for vanishing or small $J$, the repulsive interaction is reduced by the screening effect from the other orbitals but stays positive (repulsive).
On the other hand, for larger $J$,  see Fig.~\ref{fig:interaction}(c,d), $\tilde U_{1\ua\mathchar`-1\da}$ and $\tilde U_{1\ua\mathchar`-2\da}$ are not much modified, while $\tilde U_{1\ua\mathchar`-2\ua}$ is strongly reduced with increasing fluctuations and finally becomes attractive.
This behavior naturally explains the superconducting mechanism in the weak-coupling regime.

\section*{
Details of the susceptibility calculations
}

Here we explain the details how to calculate the susceptibility in the framework of DMFT [2].
For diagonal and offdiagonal orders, the relevant two-particle Green functions are given by
\begin{align}
\chi^{\rm diag}_{ij,\gm\sg\gm'\sg'} (\tau_1, \tau_2, \tau_3, \tau_4) &=
\la T_\tau c^\dg_{i\gm\sg} (\tau_1) c_{i\gm\sg} (\tau_2) c^\dg_{j\gm'\sg'} (\tau_3) c_{j\gm'\sg'} (\tau_4) \ra
-
\la T_\tau c^\dg_{i\gm\sg} (\tau_1) c_{i\gm\sg} (\tau_2) \ra \la T_\tau c^\dg_{j\gm'\sg'} (\tau_3) c_{j\gm'\sg'} (\tau_4) \ra
, \label{eqn:tpgf}
\\
\chi^{\rm offd}_{ij,\gm\sg\gm'\sg'} (\tau_1, \tau_2, \tau_3, \tau_4) &=
\la T_\tau c^\dg_{i\gm\sg} (\tau_1) c^\dg_{i\gm'\sg'} (\tau_2) c_{j\gm'\sg'} (\tau_3) c_{j\gm\sg} (\tau_4) \ra
,
\end{align}
respectively.
Here $T_\tau$ represents the imaginary time-ordering operator, and we subtract the disconnected part in Eq.~\eqref{eqn:tpgf}.
We also define the Fourier transformation with respect to time by
\begin{align}
\chi_{ij,\gm\sg\gm'\sg'}^{\al} (\imu \ep_n, \imu \ep_{n'})
=\frac{1}{\beta^2} \int_0^\beta \diff\tau_1\diff\tau_2\diff\tau_3\diff\tau_4
\, \chi^{\al}_{ij,\gm\sg\gm'\sg'} (\tau_1, \tau_2, \tau_3, \tau_4) 
\, \epn^{\imu \ep_n (\tau_2 - \tau_1) }
\, \epn^{\imu \ep_{n'} (\tau_4 - \tau_3)}
,
\end{align}
where $\al$ denotes `${\rm diag}$' or `${\rm offd}$'.
Since we are interested only in static susceptibilities, we set the bosonic frequency to zero.

Now we employ the Bethe-Salpeter equation which relates the two-particle Green function and vertex parts.
Since the vertex part $\Gamma$ is local in DMFT [2], it can be evaluated from the local two-particle Green function as
\begin{align}
\chi_{ii,\gm\sg\gm'\sg'}^{\al} (\imu \ep_n, \imu \ep_{n'})
=
\chi_{ii,\gm\sg\gm'\sg'}^{\al,0} (\imu \ep_n, \imu \ep_{n'})
+
\sum_{n_1\gm_1\sg_1}
\sum_{n_2\gm_2\sg_2}
\chi_{ii,\gm\sg\gm_1\sg_1}^{\al,0} (\imu \ep_n, \imu \ep_{n_1})
\Gamma_{i,\gm_1\sg_1\gm_2\sg_2}^{\al} (\imu \ep_{n_1}, \imu \ep_{n_2})
\chi_{ii,\gm_2\sg_2\gm'\sg'}^{\al} (\imu \ep_{n_2}, \imu \ep_{n'})
. \label{eqn:bs_eq}
\end{align}
Here the two-particle Green function without vertex parts is written as $\chi^{\al,0}$.
The site index can be neglected if the system is uniform.
The local quantities are calculated from the effective impurity system using the standard technique of CTQMC [3].
Equation~\eqref{eqn:bs_eq} may be regarded as a matrix equation with respect to the set $(n,\gm,\sg)$, and by solving this the vertex parts are obtained.
In our calculation, the number of Matsubara frequencies for this two-particle quantity is typically taken as 150.
Although the value of the susceptibility varies as a function of this cutoff, the divergent points (phase boundaries) are insensitive to it.

We note that additional contributions to the vertex parts, which cannot be written in the form of Eq.~\eqref{eqn:bs_eq}, must be considered when we have e.g. offdiagonal hybridizations.
In the present model, however, Eq.~\eqref{eqn:bs_eq} is sufficient for the derivation of the susceptibilities listed in Eq.~(4) in the main text.

The local vertex extracted above is inserted into the non-local Bethe-Salpeter equation 
\begin{align}
\chi_{ij,\gm\sg\gm'\sg'}^{\al} (\imu \ep_n, \imu \ep_{n'})
=
\chi_{ij,\gm\sg\gm'\sg'}^{\al,0} (\imu \ep_n, \imu \ep_{n'})
+
\sum_k
\sum_{n_1\gm_1\sg_1}
\sum_{n_2\gm_2\sg_2}
\chi_{ik,\gm\sg\gm_1\sg_1}^{\al,0} (\imu \ep_n, \imu \ep_{n_1})
\Gamma_{k,\gm_1\sg_1\gm_2\sg_2}^{\al} (\imu \ep_{n_1}, \imu \ep_{n_2})
\chi_{kj,\gm_2\sg_2\gm'\sg'}^{\al} (\imu \ep_{n_2}, \imu \ep_{n'}).
\end{align}
In practice, it is convenient to perform the Fourier transformation with respect to the site index before solving the matrix equation.
Thus we derive the two-particle Green function, which contain the information of the susceptibilities.
For example, the ferromagnetic susceptibility is given by
\begin{align}
\chi_{\rm FM} &= \frac{1}{N\beta} \sum_{ij}\sum_{nn'}\sum_{\gm\gm'}\sum_{\sg\sg'} 
\sg^z_{\sg\sg} \sg^z_{\sg'\sg'}
\chi^{\rm diag}_{ij,\gm\sg\gm'\sg'}(\imu\ep_n, \imu\ep_{n'})
,
\end{align}
where $\sg^z$ is a $z$-component of spin-$1/2$ Pauli matrix.
On the other hand, the inter-orbital/spin-triplet pairing susceptibility is given by
\begin{align}
\chi_{\rm SC} &= \frac{1}{N\beta} \sum_{ij}\sum_{nn'} \chi^{\rm offd}_{ij,\gm\ua\gm'\ua}(\imu\ep_n, \imu\ep_{n'}),
\end{align}
with $\gm\neq\gm'$.
The original definitions are given by Eqs.~(3) and (4) in the main text.
The other susceptiblities for e.g. orbital order can be calculated in a similar manner.

\begin{figure}[b]
\begin{center}
\includegraphics[width=65mm]{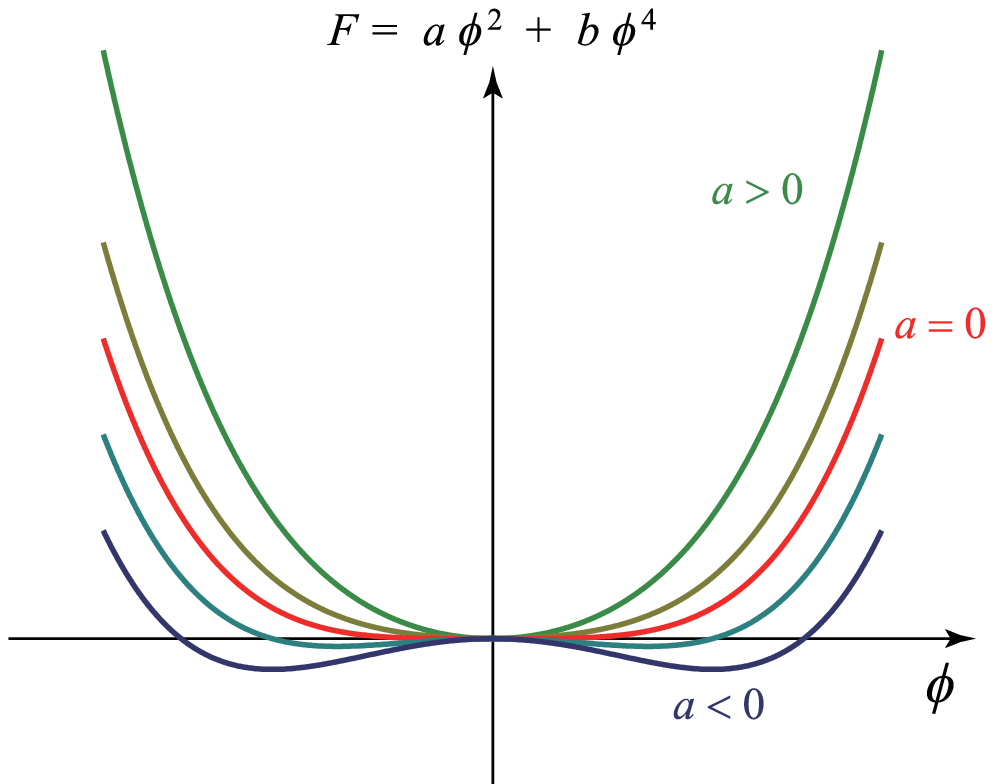}
\caption{
Landau free energy landscape above and below the transition temperature.
}
\label{fig:free_energy}
\end{center}
\end{figure}

\section*{Interpretation of the minimum in the pairing susceptibility below $T_c$}

Let us consider the meaning of the negative pairing susceptibility obtained in certain filling or interaction regimes. 
According to Landau theory, the free energy is given by 
\begin{align}
F = a \phi^2 + b \phi^4 
,
\end{align}
where $\phi$ is the pairing amplitude and $b>0$.
Since we have the relation 
\begin{align}
a = \chi_{\rm SC}^{-1} / 2
\end{align}
between the coefficient $a$ and the pairing susceptibility $\chi_{\rm SC}$, 
the sign reversal of $a$ 
signals a second-order phase transition as shown in Fig.~\ref{fig:free_energy}.
If $a < 0$, the pair amplitude is given by 
\begin{align}
\phi = \sqrt{\frac{|a|}{2b}} 
.
\end{align}
This indicates that the magnitude of the order parameter can be deduced from the value of $a$ by assuming that $b$ is a constant, even though we do not enter the symmetry broken phase.
More specifically, the minimum of the (negative) inverse susceptibility $\chi_{\rm SC}^{-1}$ coincides with the maximum of the pairing amplitude.
Although this argument is valid only near the transition temperature, we expect that the 
connection 
between minimum $\chi_{\rm SC}^{-1}$ and maximum pairing amplitude approximately holds in a wider temperature range.

\section*{
Additional data for the phase diagram
}

\begin{figure}[h]
\begin{center}
\includegraphics[width=170mm]{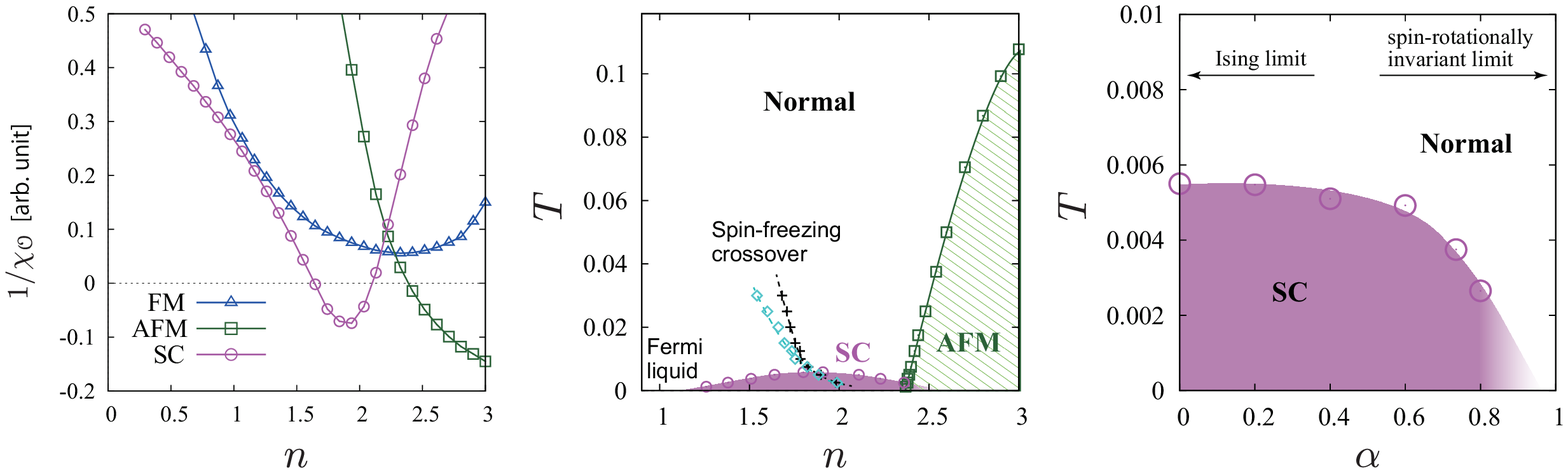}
\caption{
Results for $U=1$:
(Left) filling dependence of susceptibilities at $T=0.005$, (Middle) filling-temperature phase diagram, and (Right) spin anisotropy ($\alpha$) dependence of the 
superconducting transition temperature
at $n=2$.
}
\label{fig:phase_suppl}
\end{center}
\end{figure}

In the main text, we discuss the superconductivity by choosing several interaction parameters.
Here we show that the behavior remains qualitatively the same when we change the parameters.
Figure~\ref{fig:phase_suppl} shows the filling and anisotropy parameter $\alpha$ dependences of the inverse susceptibilities and phase diagrams at $U=1$.
The results are qualitatively similar to those for $U=0.75$ and $U=0.875$ shown in the main text.

\vspace{3mm}
{\bf References}

{[1]} K. Inaba and S. Suga, Phys. Rev. Lett. {\bf 108}, 255301 (2012).

{[2]} A. Georges, G. Kotliar, W. Krauth and M. J. Rozenberg, Rev. Mod. Phys. {\bf 68}, 13 (1996).

{[3]} E. Gull, A. J. Millis, A. I. Lichtenstein, A. N. Rubtsov, M. Troyer, and P. Werner, Rev. Mod. Phys. {\bf 83}, 349 (2011).


\begin{thebibliography}{99}
\bibitem{footnote_soc}
Although an explicit treatment of the spin-orbit coupling is necessary for a quantitative analysis, both approaches generate an anisotropy in spin space and are expected to yield similar results.
\end{thebibliography}
\end{document}